\documentclass[acmtocl]{acmtrans2m}

\usepackage{amsmath}
\usepackage{amssymb}
\usepackage{mathrsfs}  
\usepackage{graphicx}
\usepackage{color}
\usepackage[all]{xy}

\newtheorem{theorem}{Theorem}[section]
\newtheorem{conjecture}[theorem]{Conjecture}
\newtheorem{corollary}[theorem]{Corollary}

\newtheorem{lemma}[theorem]{Lemma}
\newdef{definition}[theorem]{Definition}
\newdef{remark}[theorem]{Remark}
\newdef{remarks}[theorem]{Remarks}

\newcommand{\CSP}[0]{\ensuremath{\textsc{CSP}}}

\newcommand{\QCSP}[0]{\ensuremath{\textsc{QCSP}}}

\newcommand{\Aut}[0]{\ensuremath{\mathsf{Aut}}}

\newcommand{\Inv}[0]{\ensuremath{\mathsf{Inv}}}
\newcommand{\shE}[0]{\ensuremath{\mathsf{shE}}}

\newcommand{\NP}[0]{\ensuremath{\mathsf{NP}}}

\newcommand{\coNP}[0]{\ensuremath{\mathsf{co\mbox{-}NP}}}

\newcommand{\Logspace}[0]{\ensuremath{\mathsf{L}}}

\newcommand{\Ptime}[0]{\ensuremath{\mathsf{P}}}
\newcommand{\Pspace}[0]{\ensuremath{\mathsf{Pspace}}}

\newcommand{\FO}[0]{\ensuremath{\mathsf{FO}}}

\newcommand{\mylogic}{\ensuremath{\{\exists, \forall, \wedge,\vee \} \mbox{-}\mathsf{FO}}}
\newcommand{\posFO}{\ensuremath{\{\exists, \forall, \wedge,\vee,= \} \mbox{-}\mathsf{FO}}}
\newcommand{\tuple}[1]{\ensuremath{\mathbf{#1}}}

\newcommand{\notmodels}{\ensuremath{ \models \hspace{-3mm} / \hspace{2mm} }}

\newcommand{\she}[3]{
\resizebox{!}{.4cm}{
\ensuremath{
\begin{array}{c|c}
0 & #1 \\
\hline
1 & #2 \\
\hline
2 & #3 
\end{array}
}
}
}

\newcommand{\shefour}[4]{
\resizebox{!}{.5cm}{
\ensuremath{
\begin{array}{c|c}
0 & #1 \\
\hline
1 & #2 \\
\hline
2 & #3 \\
\hline
3 & #4 
\end{array}
}
}
}

\newcommand{\shee}[2]{
\resizebox{!}{.4cm}{
\ensuremath{
\begin{array}{c|c}
0 & #1 \\
\hline
1 & #2 \\
\end{array}
}
}
}

\newcommand{\BibTeX}{{\rm B\kern-.05em{\sc i\kern-.025em b}\kern-.08em
    T\kern-.1667em\lower.7ex\hbox{E}\kern-.125emX}}

\title{The complexity of positive first-order logic without equality}
\author{FLORENT MADELAINE\\Univ Clermont1, EA2146.
\and
BARNABY MARTIN\\Durham University.}
\begin{abstract}
We study the complexity of evaluating positive equality-free sentences of first-order (FO) logic over a fixed, finite structure $\mathcal{B}$. This may be seen as a natural generalisation of the non-uniform quantified constraint satisfaction problem $\QCSP(\mathcal{B})$. We introduce surjective hyper-endomorphisms and use them in proving a Galois connection that characterises definability in positive equality-free FO. Through an algebraic method, we derive a complete complexity classification for our problems as $\mathcal{B}$ ranges over structures of size at most three. Specifically, each problem is either in \Logspace, is \NP-complete, is \coNP-complete or is \Pspace-complete.
\end{abstract}




\begin{document}

\setcounter{page}{111}

\begin{bottomstuff}
Author's addresses: Florent Madelaine, Univ Clermont1, EA2146, Laboratoire d'algorithmique et d'image de Clermont-Ferrand, Aubi\`ere, F-63170, France. \texttt{florent.madelaine@u-clermont1.fr}. Barnaby Martin, School of Engineering and Computing Sciences, Durham University, Durham DH1 3LE, U.K. \texttt{barnabymartin@gmail.com}. 
\end{bottomstuff}
\maketitle

\section{Introduction}
The evaluation problem under a logic $\mathcal{L}$ -- here always a fragment of first-order logic (\FO) -- takes as input a structure (model) $\mathcal{B}$ and a sentence $\varphi$ of $\mathcal{L}$, and asks whether $\mathcal{B} \models \varphi$.\footnote{We resist the better known terminology of `model checking problem' because in the majority of this paper we consider the structure $\mathcal{B}$ to be fixed.} When $\mathcal{L}$ is the \emph{existential conjunctive positive} fragment of \FO, $\{ \exists, \wedge \}$-\FO, the evaluation problem is equivalent to the much-studied \emph{constraint satisfaction problem} (\CSP). Similarly, when $\mathcal{L}$ is the \emph{(quantified) conjunctive positive} fragment of \FO, $\{ \exists, \forall, \wedge \}$-\FO, the evaluation problem is equivalent to the well-studied \emph{quantified constraint satisfaction problem} (\QCSP).
In this manner, the \QCSP\ is the generalisation of the \CSP\ in which universal quantification is restored to the mix. In both cases it is essentially irrelevant whether or not equality is permitted in the sentences, as it may be propagated out by substitution. Much work has been done on the parameterisation of these problems by the structure $\mathcal{B}$ -- that is, where $\mathcal{B}$ is fixed and only the sentence is input. It is conjectured \cite{FederVardi} that the ensuing problems $\CSP(\mathcal{B})$ attain only the complexities \Ptime\ and \NP-complete. This may appear surprising given that 1.) so many natural \NP\ problems may be expressed as \CSP s (see, e.g., myriad examples in \cite{jeavons98algebraic}) and 2.) \NP\ itself does not have this `dichotomy' property (assuming $\Ptime \neq \NP$) \cite{Ladner}. While this \emph{dichotomy conjecture} remains open, it has been proved for certain classes of $\mathcal{B}$ (e.g., for structures of size at most three \cite{BulatovJACM} and for undirected graphs \cite{HellNesetril}). 
The like parameterisation of the \QCSP\ is also well-studied, and while no overarching polychotomy has been conjectured, only the complexities \Ptime, \NP-complete and \Pspace-complete are known to be attainable (for trichotomy results on certain classes see \cite{OxfordQuantifiedConstraints,DBLP:conf/cie/MartinM06}, as well as the dichotomy for boolean structures, e.g., in \cite{Nadia}).

In previous work, \cite{CiE2008,DBLP:journals/corr/abs-cs-0609022}, we have studied the evaluation problem, parameterised by the structure, under various fragments of \FO\ obtained by restrictions on which of the symbols of $\{\exists,\forall,\wedge,\vee,\neg,=,\neq\}$ is permitted. 
Of course, many of the ostensibly $2^7$ such fragments may be discarded as totally trivial or as repetitions through de Morgan duality. There are four fragments each equivalent to the \CSP\ and \QCSP: these are $\{ \exists, \wedge \}$-\FO, $\{ \exists,\wedge, = \}$-\FO, $\{ \forall, \vee \}$-\FO, $\{ \forall, \vee, \neq \}$-\FO\ and $\{ \exists, \forall, \wedge \}$-\FO, $\{ \exists, \forall, \wedge, = \}$-\FO, $\{ \exists, \forall, \vee \}$-\FO, $\{ \exists, \forall, \vee, \neq \}$-\FO, respectively. 
Here, equivalent means that a complexity classification for one yields a complexity classification for the other; but, the complexity classes need not be the same. For example, the class of problems given by fixing the structure under $\{ \exists, \wedge \}$-\FO\ would display dichotomy between \Ptime\ and \NP-complete iff the like class of problems under $\{ \forall, \vee \}$-\FO\ displays dichotomy between \Ptime\ and \coNP-complete. 
Various complexity classifications are obtained in \cite{CiE2008,DBLP:journals/corr/abs-cs-0609022} and it is observed that the only interesting fragment, other than the eight associated with \CSP\ and \QCSP, is $\{ \exists, \forall, \wedge, \vee \}$-\FO.\footnote{
For many of the other fragments the complexity classification is nearly trivial. For example, this is true for $\{ \exists, \wedge, \vee \}$-\FO, $\{ \forall, \wedge, \vee \}$-\FO\ and $\{ \exists, \forall, \wedge, \vee, \neg \}$-\FO\ (also for these classes with $=$ or $\neq$). For others the classification may be read through the Schaefer classification for boolean \CSP\ and \QCSP, because computational hardness is clear over fixed structures of size at least three. For example, this is the case for $\{ \exists, \wedge, \neq \}$-\FO, $\{ \forall, \vee, = \}$-\FO\ and $\{ \exists, \forall, \wedge, \neq \}$-\FO, $\{ \exists, \forall, \vee, = \}$-\FO. Note that the consideration of $\neq$  is not explicit in \cite{CiE2008,DBLP:journals/corr/abs-cs-0609022}. Similarly, fragments involving both quantifiers and $=$ or $\neq$ are not explicitly considered. In both cases, the results may be read off from de Morgan duality together with standard Schaefer class results (for which we refer to \cite{Nadia}).
}
The evaluation problem over $\{ \exists, \forall, \wedge, \vee \}$-\FO\ may be seen as the generalisation of the \QCSP\ in which disjunction is returned to the mix. Note that the absence of equality is here important, as there is no general method for its being propagated out by substitution. Indeed, we will see that evaluating the related fragment $\{ \exists, \forall, \wedge, \vee, = \}$-\FO\ is \Pspace-complete on any structure $\mathcal{B}$ of size at least two. 

In this paper we initiate a study of the evaluation problem for the fragment $\{ \exists, \forall, \wedge, \vee \}$-\FO\ over a fixed relational $\mathcal{B}$ -- the problem we denote $\mylogic(\mathcal{B})$. We demonstrate at least that this class displays a complexity-theoretic richness absent from those other fragments that are not associated with the \CSP\ or \QCSP. It is possibly to be hoped, however, that a full classification for this class is not as resistant as that for the \CSP\ or \QCSP. We undertake our study through the algebraic method that has been so fruitful in the study of the \CSP\ and \QCSP\ (see \cite{JeavonsCohenGyssonsJACM,BulatovJACM,OxfordQuantifiedConstraints,HubieSicomp}). To this end, we define surjective hyper-endomorphisms and use them to define a new Galois connection that characterises definability under $\{ \exists, \forall, \wedge, \vee \}$-\FO.\footnote{While this Galois connection appears here for the first time, it does follow a general recipe as outlined, e.g., in \cite{FerdinandDagstuhl}. Note that it is not clear that the many different Galois connections associated with fragments of \FO\ can be proved in a straightforwardly uniform manner.} We are able to prove a complete complexity classification for $\mylogic(\mathcal{B})$ when $\mathcal{B}$ ranges over structures of size at most three. On the class of boolean structures we see dichotomy between \Logspace\ and \Pspace-complete. 
On the class of structures of size three we see tetrachotomy between \Logspace, \NP-complete, \coNP-complete and \Pspace-complete. Some of the results that appear in this paper had been obtained through adhoc methods in \cite{DBLP:journals/corr/abs-0808-0647} -- although there the tetrachotomy extends only to digraphs and not arbitrary relational structures. Also, little insight was provided as to the underlying properties of the classification. It is a pleasing consequence of our algebraic approach that we can give quite simple explanation to the delineation of our subclasses.

The paper is organised as follows. In Section~\ref{sec:preliminaries}, we introduce the preliminaries, including the relevant Galois connection together with the central notions of surjective hyper-endomorphism (she) and down-she-monoid. In Section~\ref{sec:classification-methods}, we outline conditions under which the problem $\mylogic(\mathcal{B})$ either drops from or attains maximal complexity. In Section~\ref{sec:classification-results} we classify the complexity of the problems $\mylogic(\mathcal{B})$, when $\mathcal{B}$ ranges over, firstly, boolean structures and, secondly, structures of size three. In the first instance a dichotomy -- between \Logspace\ and \Pspace-complete -- is obtained; in the second instance a tetrachotomy -- between \Logspace, \NP-complete, \coNP-complete and \Pspace-complete -- is obtained. We conclude, in Section~\ref{sec:final-remarks}, with some final remarks. 

An extended abstract of this paper has appeared as \cite{LICS2009}.

\section{Preliminaries}
\label{sec:preliminaries}

Throughout, let $\mathcal{B}$ be a finite structure, with domain $B$, over the finite relational signature $\sigma$. Let \mylogic\ and \posFO\ be the positive fragments of first-order (FO) logic, without and with equality, respectively. An \emph{extensional} relation is one that appears in the signature $\sigma$. We will usually denote extensional relations of $\mathcal{B}$ by $R$ and other relations by $S$ (or by some formula that defines them). In \mylogic\, the atomic formulae are exactly substitution instances of extensional relations. The problem $\mylogic(\mathcal{B})$ has:
\begin{itemize}
\item Input: a sentence $\varphi \in \mylogic$.
\item Question: does $\mathcal{B} \models \varphi?$
\end{itemize}
The related problem $\posFO(\mathcal{B})$ permits sentences $\varphi$ that may involve equalities, in the obvious way.
When $\mathcal{B}$ is of size one, the evaluation of any \FO\ sentence may be accomplished in \Logspace\ (essentially, the quantifiers are irrelevant and the problem amounts to the \emph{boolean sentence value problem}, see \cite{NLynch}). In this case, it follows that both $\mylogic(\mathcal{B})$ and $\posFO(\mathcal{B})$ are also in \Logspace.

Consider the set $B$ and its power set $\mathfrak{P}(B)$. A \emph{hyper-operation} on $B$ is a function $f$ from $B$ to $\mathfrak{P}(B) \setminus \{\emptyset\}$ (that the image may not be the empty set corresponds to the hyper-operation being \emph{total}, in the parlance of \cite{BornerTotalMultifunctions}). If the hyper-operation $f$ has the additional property that
\begin{itemize}
\item for all $y \in B$, there exists $x \in B$ such that $y \in f(x)$,
\end{itemize}
then we designate (somewhat abusing terminology) $f$ \emph{surjective}. A surjective hyper-operation (shop) in which each element is mapped to a singleton set is identified with a \emph{permutation} (bijection). A \emph{surjective hyper-endomorphism} (she) of $\mathcal{B}$ is a surjective hyper-operation $f$ on $B$ that satisfies, for all extensional relations $R$ of $\mathcal{B}$,
\begin{itemize}
\item if $\mathcal{B} \models R(x_1,\ldots,x_i)$ then, for all $y_1 \in f(x_1),\ldots,y_i \in f(x_i)$, $\mathcal{B} \models R(y_1,\ldots,y_i)$.
\end{itemize}
More generally, for $r_1,\ldots,r_k \in B$, we say $f$ is \emph{a she from} $(\mathcal{B},r_1,\ldots,r_k)$ to $(\mathcal{B},r'_1,\ldots$ $,r'_k)$ if $f$ is a she of $\mathcal{B}$ and $r'_1 \in f(r_1), \ldots, r'_k \in f(r_k)$.
A she may be identified with a \emph{surjective endomorphism} if each element is mapped to a singleton set. On finite structures surjective endomorphisms are necessarily automorphisms.

For $b_1,\ldots,b_{|B|}$ an enumeration of the elements of $\mathcal{B}$, let the quantifier-free formula $\Phi_\mathcal{B}(v_1,\ldots,v_{|B|})$ be a conjunction of the positive facts of $\mathcal{B}$, where the variables $v_1,\ldots,v_{|B|}$ correspond to the elements $b_1,\ldots,b_{|B|}$. That is, for $R$ an extensional relation of $\mathcal{B}$, $R(v_{\lambda_1},\ldots,v_{\lambda_i})$ appears as an atom in $\Phi_\mathcal{B}$ iff $\mathcal{B} \models R(b_{\lambda_1},\ldots,b_{\lambda_i})$. For example, let $\mathcal{K}_3$ be the antireflexive $3$-clique, that is the structure with domain $\{0,1,2\}$ and single binary relation
\[ E:=\{(0,1),(1,0),(1,2),(2,1),(2,0),(0,2)\}.\]
Then 
\[ \Phi_{\mathcal{K}_3}(v_0,v_1,v_2):= E(v_0,v_1) \wedge E(v_1,v_0) \wedge E(v_1,v_2) \wedge E(v_2,v_1) \wedge E(v_2,v_0) \wedge E(v_0,v_2).\]
The existential sentence $\exists v_1,\ldots,v_{|B|} \ \Phi_\mathcal{B}(v_1,\ldots,v_{|B|})$ is known as the \emph{canonical query} of $\mathcal{B}$. More generally, for a (not necessarily distinct) $l$-tuple of elements $\tuple{r}:=(r_1,\ldots,r_l) \in B^l$, define the quantifier-free $\Phi_{\mathcal{B}(\tuple{r})}(v_1,\ldots,v_l)$ to be the conjunction of the positive facts of $\tuple{r}$, where the variables $v_1,\ldots,v_l$ correspond to the elements $r_1,\ldots,r_l$. That is, $R(v_{\lambda_1},\ldots,v_{\lambda_i})$ appears as an atom in $\Phi_{\mathcal{B}(\tuple{r})}$ iff $\mathcal{B} \models R(r_{\lambda_1},\ldots,r_{\lambda_i})$. For example, 
\[ \Phi_{\mathcal{K}_3(0,0,2)}(v_0,v_1,v_2):= E(v_0,v_2) \wedge E(v_2,v_0) \wedge E(v_1,v_2) \wedge E(v_2,v_1).\]

We refer to elements in $\mathcal{B}$ as $r,s,t$ (also $x,y$), or $b_1,\ldots,b_{|B|}$ when this is an enumeration. We reserve $u,v,w$ to refer to variables in \FO\ formulae.

\subsection{Galois Connections}

For a set $F$ of shops on the finite domain $B$, let $\Inv(F)$ be the set of relations on $B$ of which each $f \in F$ is a she (when these relations are viewed as a structure over $B$). We say that $S \in \Inv(F)$ is invariant or \emph{preserved} by (the shops in) $F$. Let $\shE(\mathcal{B})$ be the set of shes of $\mathcal{B}$. Let $\Aut(\mathcal{B})$ be the set of automorphisms of $\mathcal{B}$.

Let $\langle \mathcal{B} \rangle_{\mylogic}$ and $\langle \mathcal{B} \rangle_{\posFO}$ be the sets of relations that may be defined on $\mathcal{B}$ in \mylogic\ and \posFO, respectively.

\begin{lemma}
\label{lemma:galois-connection-by-types}
Let $\tuple{r}:=(r_1,\ldots,r_k)$ be a $k$-tuple of elements of $\mathcal{B}$. There exists:
\begin{itemize}
\item[$(i).$] a formula $\theta_\tuple{r}(u_1,\ldots,u_k) \in \posFO$ s.t. $(\mathcal{B}, r'_1,\ldots, r'_k) \models \theta_\tuple{r}(u_1,\ldots,u_k)$ iff there is an automorphism from $(\mathcal{B}, r_1,\ldots, r_k)$ to $(\mathcal{B}, r'_1,\ldots, r'_k)$.
\item[$(ii).$] a formula $\theta_\tuple{r}(u_1,\ldots,u_k) \in \mylogic$ s.t. $(\mathcal{B}, r'_1,\ldots, r'_k) \models \theta_\tuple{r}(u_1,\ldots,u_k)$ iff there is a she from $(\mathcal{B}, r_1,\ldots, r_k)$ to $(\mathcal{B}, r'_1,\ldots, r'_k)$.
\end{itemize}
\end{lemma}
\begin{proof}
For Part $(i)$, let $b_1,\ldots,b_{|B|}$ an enumeration of the elements of $\mathcal{B}$ and $\Phi_\mathcal{B}(v_1,\ldots,v_{|B|})$ be the associated conjunction of positive facts. Set $\theta_\tuple{r}(u_1,\ldots,u_k):=$
\[ \exists v_1,\ldots,v_{|B|} \ \Phi_\mathcal{B}(v_1,\ldots,v_{|B|}) \wedge \forall v \ (v=v_1 \vee \ldots \vee v=v_{|B|}) \wedge u_1=v_{\lambda_1} \wedge \ldots \wedge u_k=v_{\lambda_k}, \]
where $r_1 = b_{\lambda_1}$, \ldots, $r_k = b_{\lambda_k}$. The forward direction follows since $\mathcal{B}$ is finite, so any surjective endomorphism is necessarily an automorphism. The backward direction follows since all first-order formulae are preserved by automorphism.


[Part $(ii)$.]
This will require greater dexterity. Let $\tuple{r} \in B^k$, $\tuple{s} := (b_1,\ldots,b_{|B|})$ be an enumeration of $B$ and $\tuple{t} \in B^{|B|}$. Recall that $\Phi_{\mathcal{B}(\tuple{r},\tuple{s})}(u_1,\ldots,u_k,v_1,\ldots,v_{|B|})$ is a conjunction of the positive facts of $(\tuple{r},\tuple{s})$, where the variables $(\tuple{u},\tuple{v})$ correspond to the elements $(\tuple{r},\tuple{s})$. Similarly, $\Phi_{\mathcal{B}(\tuple{r},\tuple{s},\tuple{t})}(u_1,\ldots,u_k,v_1,\ldots,v_{|B|},w_1,\ldots,w_{|B|})$ is the conjunction of the positive facts of $(\tuple{r},\tuple{s},\tuple{t})$, where the variables $(\tuple{u},\tuple{v},\tuple{w})$ correspond to the elements $(\tuple{r},\tuple{s},\tuple{t})$.
Set $\theta_\tuple{r}(u_1,\ldots,u_k):=$
\[
\exists v_1,\ldots,v_{|B|} \ \Phi_{\mathcal{B}(\tuple{r},\tuple{s})}(u_1,\ldots,u_k,v_1,\ldots,v_{|B|}) \wedge \forall w_1 \ldots w_{|B|} \ \ \ \ \ \ \ \ \ \ \ \ \ \ \ \ \ \ \ \ \ \ \ \ \ \ \ \ \ \ \ \ \ \ \ \ \ \ \ \
\]
\[
\ \ \ \ \ \ \ \ \ \ \ \ \ \ \ \ \ \ \ \ \ \ \ \ \ \ \ \ \ \ \ \ \ \ \ \ \ \ \ \ \ \ \ \bigvee_{\tuple{t} \in B^{|B|}} \Phi_{\mathcal{B}(\tuple{r},\tuple{s},\tuple{t})}(u_1,\ldots,u_k,v_1,\ldots,v_{|B|},w_1,\ldots,w_{|B|}).
\]

[Part $(ii)$, backwards.] Suppose $f$ is a she from $(\mathcal{B}, r_1,\ldots, r_k)$ to $(\mathcal{B}', r'_1,\ldots, r'_k)$, where $\mathcal{B}':=\mathcal{B}$ (we will wish to differentiate the two occurrences of $\mathcal{B}$). We aim to prove that $\mathcal{B}'  \models \theta_\tuple{r}(r'_1,\ldots, r'_k)$. Choose arbitrary $s'_1 \in f(b_1),\ldots,s'_{|B|} \in f(b_{|B|})$ as witnesses for $v_1,\ldots,v_{|B|}$. Let $\tuple{t}':=(t'_1,\ldots,t'_{|B|}) \in B'^{|B|}$ be any valuation of $w_1,\ldots,w_{|B|}$ and take arbitrary $t_1,\ldots,t_{|B|}$ s.t. $t'_1 \in f(t_1)$, \ldots, $t'_{|B|} \in f(t_{|B|})$ (here we use surjectivity). Let $\tuple{t}:=(t_1,\ldots,t_{|B|})$. It follows from the definition of she that
\[ \mathcal{B}' \ \models \ \Phi_{\mathcal{B}(\tuple{r},\tuple{s})}(r'_1,\ldots,r'_k,s'_1,\ldots,s'_{|B|}) \wedge \Phi_{\mathcal{B}(\tuple{r},\tuple{s},\tuple{t})}(r'_1,\ldots,r'_k,s'_1,\ldots,s'_{|B|},t'_1,\ldots,t'_{|B|}). \]

[Part $(ii)$, forwards.] Assume that $\mathcal{B}' \models \theta_\tuple{r}(r'_1,\ldots, r'_k)$, where $\mathcal{B}':=\mathcal{B}$. Let $b'_1, \ldots,b'_{|B|}$ be an enumeration of $B':=B$.\footnote{One may imagine $b_1, \ldots,b_{|B|}$ and $b'_1, \ldots,b'_{|B|}$ to be the same enumeration, but this is not essential. In any case, we will wish to keep the dashes on the latter set to remind us they are in $\mathcal{B}'$ and not $\mathcal{B}$.} Choose some witness elements $s'_1,\ldots,s'_{|B|}$ for $v_1,\ldots,v_{|B|}$ and a witness tuple $\tuple{t}:=$ $(t_1,\ldots,t_{|B|}) \in B^{|B|}$ s.t.
\[ (\dagger) \ \mathcal{B}' \models \ \Phi_{\mathcal{B}(\tuple{r},\tuple{s})}(r'_1,\ldots,r'_k,s'_1,\ldots,s'_{|B|}) \wedge \Phi_{\mathcal{B}(\tuple{r},\tuple{s},\tuple{t})}(r'_1,\ldots,r'_k,s'_1,\ldots,s'_{|B|},b'_1,\ldots,b'_{|B|}). \]
Consider the following partial hyper-operations from $B \rightarrow \mathfrak{P}(B') \setminus \{\emptyset\}$.
\begin{itemize}
\item[1.] $f_{\tuple{r}}$ given by $f_{\tuple{r}}(r_i):= \{ r'_i \}$, for $1 \leq i \leq k$.
\item[2.] $f_{\tuple{s}}$ given by $f_{\tuple{s}}(b_i)=\{ s'_i \}$, for $1 \leq i \leq |B|$. \ \ \ \ \ \ \ \ (totality.)
\item[3.] $f_{\tuple{t}}$ given by $b'_i \in f_{\tuple{t}}(b_j)$ iff $t_i=b_j$, for $1 \leq i,j \leq |B|$. \ \ \ \ \ \ \ \ (surjectivity.) 
\end{itemize}
Let $f:=f_{\tuple{r}} \cup f_{\tuple{s}} \cup f_{\tuple{t}}$; $f$ is a hyper-operation whose surjectivity is guaranteed by $f_{\tuple{t}}$ (note that totality is guaranteed by $f_{\tuple{s}}$). That $f$ is a she follows from the right-hand conjunct of $(\dagger)$.
\end{proof}
\begin{theorem}
\label{thm:galois-connection}
For a finite structure $\mathcal{B}$ we have
\begin{itemize}
\item[$(i).$] $\langle \mathcal{B} \rangle_{\posFO} = \Inv(\Aut(\mathcal{B}))$ and
\item[$(ii).$] $\langle \mathcal{B} \rangle_{\mylogic} = \Inv(\shE(\mathcal{B}))$.
\end{itemize}
\end{theorem}
\begin{proof}
Part $(i)$ is well-known and may be proved in a similar, albeit simpler, manner to Part $(ii)$, which we now prove.

[$\varphi(\tuple{v}) \in \langle \mathcal{B} \rangle_{\mylogic} \ \Rightarrow \ \varphi(\tuple{v}) \in \Inv(\shE(\mathcal{B}))$.] This is proved by induction on the complexity of $\varphi(\tuple{v})$.

(Base Case.) $\varphi(\tuple{v}):=R(\tuple{v})$.\footnote{The variables $\tuple{v}$ may appear multiply in $R$ and in any order. Thus $R$ is an instance of an extensional relation under substitution and permutation of positions.} Follows from the definition of she.

(Inductive Step.) There are four subcases. We progress through them in a workmanlike fashion. Take $f \in \shE(\mathcal{B})$.

$\varphi(\tuple{v}):=\psi(\tuple{v}) \wedge \psi'(\tuple{v})$.\footnote{The presence of, e.g., $\tuple{v}$ in $\psi(\tuple{v}) \wedge \psi'(\tuple{v})$ should not be taken as indication that \emph{all} $\tuple{v}$ appear free in both $\psi$ and $\psi'$.} Let $\tuple{v}:=(v_1,\ldots,v_l)$. Suppose $\mathcal{B} \models \varphi(x_1,\ldots,x_l)$; then both $\mathcal{B} \models \psi(x_1,\ldots,x_l)$ and $\mathcal{B} \models \psi'(x_1,\ldots,x_l)$. By Inductive Hypothesis (IH), for any $y_1\in f(x_1),\ldots,y_l\in f(x_l)$, both $\mathcal{B} \models \psi(y_1,\ldots,y_l)$ and $\mathcal{B} \models \psi'(y_1,\ldots,y_l)$, whence $\mathcal{B} \models \varphi(y_1,\ldots,y_l)$.

$\varphi(\tuple{v}):=\psi(\tuple{v}) \vee \psi'(\tuple{v})$. Let $\tuple{v}:=(v_1,\ldots,v_l)$. Suppose $\mathcal{B} \models \varphi(x_1,\ldots,x_l)$; then one of $\mathcal{B} \models \psi(x_1,\ldots,x_l)$ or $\mathcal{B} \models \psi'(x_1,\ldots,x_l)$; w.l.o.g. the former. By IH, for any $y_1\in f(x_1),\ldots,y_l\in f(x_l)$, $\mathcal{B} \models \psi(y_1,\ldots,y_l)$, whence $\mathcal{B} \models \varphi(y_1,\ldots,y_l)$.

$\varphi(\tuple{v}):=\forall w \ \psi(\tuple{v},w)$. Let $\tuple{v}:=(v_1,\ldots,v_l)$. Suppose $\mathcal{B} \models \forall w \ \psi(x_1,\ldots,x_l,w)$; then for each $x'$, $\mathcal{B} \models \psi(x_1,\ldots,x_l,x')$. By IH, for any $y_1\in f(x_1),\ldots,y_l\in f(x_l)$, we have for all $y'$ (remember $f$ is surjective), $\mathcal{B} \models \psi(y_1,\ldots,y_l,y')$, whereupon $\mathcal{B} \models \forall w \ \psi(y_1,\ldots,y_l,w)$.

$\varphi(\tuple{v}):=\exists w \ \psi(\tuple{v},w)$. Let $\tuple{v}:=(v_1,\ldots,v_l)$. Suppose $\mathcal{B} \models \exists w \ \psi(x_1,\ldots,x_l,w)$; then for some $x'$, $\mathcal{B} \models \psi(x_1,\ldots,x_l,x')$. By IH, for any $y_1\in f(x_1),\ldots,y_l\in f(x_l),y' \in f(x')$ (remember $f(x')$ can not be empty), $\mathcal{B} \models \psi(y_1,\ldots,y_l,y')$, whereupon $\mathcal{B} \models \exists w \ \psi(y_1,\ldots,y_l,w)$.

[$S \in \Inv(\shE(\mathcal{B})) \ \Rightarrow \ S \in \langle \mathcal{B} \rangle_{\mylogic}.$] Consider the $k$-ary relation $S \in \Inv(\shE(\mathcal{B}))$. Let $\tuple{r}_1,\ldots,\tuple{r}_m$ be the tuples of $S$. Set
\[ \theta_S(u_1,\ldots,u_k) \ := \ \theta_{\tuple{r}_1}(u_1,\ldots,u_k) \vee \ldots \vee \theta_{\tuple{r}_m}(u_1,\ldots,u_k). \]
Manifestly, $\theta_S(u_1,\ldots,u_k) \in \mylogic$. For $\tuple{r}_i:=(r_{i1},\ldots,r_{ik})$, note that $(\mathcal{B},r_{i1},\ldots,r_{ik}) \models \theta_{\tuple{r}_i}(u_1,\ldots,u_k)$ (the `identity' she will be formally introduced in the next section). That $\theta_S(u_1,\ldots,u_k)=S$ now follows from Part $(ii)$ of Lemma~\ref{lemma:galois-connection-by-types}, since $S \in \Inv(\shE(\mathcal{B}))$.
\end{proof}
Let $\leq_{\Logspace}$ indicate the existence of a logspace many-to-one reduction. The following theorem is our counterpart to Corollary 4.11 of \cite{jeavons98algebraic} (for CSP) and Theorem 3.1 of \cite{OxfordQuantifiedConstraints} (for QCSP).
\begin{theorem}
\label{thm:she-reduction}
Let $\mathcal{B}$ and $\mathcal{B}'$ be finite structures over the same domain $B$.
\begin{itemize}
\item[$(i).$] If $\Aut(\mathcal{B}) \subseteq \Aut(\mathcal{B}')$ then $\posFO(\mathcal{B}') \leq_{\Logspace} \posFO(\mathcal{B})$.
\item[$(ii).$] If $\shE(\mathcal{B}) \subseteq \shE(\mathcal{B}')$ then $\mylogic(\mathcal{B}') \leq_{\Logspace} \mylogic(\mathcal{B})$.
\end{itemize}
\end{theorem}
\begin{proof}
Again, Part $(i)$ is well-known and the proof is similar to that of Part $(ii)$, which we give. If $\shE(\mathcal{B}) \subseteq \shE(\mathcal{B}')$, then $\Inv(\shE(\mathcal{B}')) \subseteq \Inv(\shE(\mathcal{B}))$. From Theorem~\ref{thm:galois-connection}, it follows that $\langle \mathcal{B}' \rangle_{\mylogic}$ $\subseteq \langle \mathcal{B} \rangle_{\mylogic}$. Recalling that $\mathcal{B}'$ contains only a finite number of extensional relations, we may therefore effect a Logspace reduction from $\mylogic(\mathcal{B}')$ to $\mylogic(\mathcal{B})$ by straightforward substitution of predicates.
\end{proof}

\subsection{Down-she-monoids}

Consider a finite domain $B$. The \emph{identity} shop $id_B$ is defined by $x \mapsto \{x\}$. Given shops $f$ and $g$, define the \emph{composition} $g \circ f$ by $x \mapsto \{ z : \exists y \ z \in g(y) \wedge y \in f(x) \}$. Finally, a shop $f$ is a \emph{sub-shop} of $g$ -- denoted $f \subseteq g$ -- if $f(x) \subseteq g(x)$, for all $x$.  
A set of surjective shops on a finite set $B$ is a \emph{down-she-monoid} (DSM), if it contains $id_B$, and is closed under composition and sub-shops (of course, not all sub-hyper-operations of a shop are surjective -- we are only concerned with those that are). $id_B$ is a she of all structures, and, if $f$ and $g$ are shes of $\mathcal{B}$, then so is $g \circ f$. Further, if $g$ is a she of $\mathcal{B}$, then so is $f$ for all (surjective) $f \subseteq g$. It follows that $\shE(\mathcal{B})$ is always a DSM. The DSMs of $B$ form a lattice under (set-theoretic) inclusion and, as per the Galois connection of the previous section, classify the complexities of $\mylogic(\mathcal{B})$. If $F$ is a set of shops on $B$, then let $\langle F \rangle$ denote the minimal DSM containing the operations of $F$. If $F$ is the singleton $\{f\}$,  then, by abuse of notation, we write $\langle f \rangle$ instead of $\langle \{f\} \rangle$ 

For a shop $f$, define its inverse $f^{-1}$ by $x \mapsto \{y : x \in f(y)\}$. Note that $f^{-1}$ is also a shop and $(f^{-1})^{-1}=f$, though $f \circ f^{-1}=id_B$ only if $f$ is a permutation. For a set of shops $F$, let $F^{-1}:=\{f^{-1}:f \in F\}$. If $F$ is a DSM then so is $F^{-1}$. We will see this algebraic duality resonates with the de Morgan duality of $\exists$ and $\forall$, and the complexity-theoretic duality of $\NP$ and $\coNP$. However, we resist discussing it further as it plays no direct role in the derivation of our results.

A \emph{permutation subgroup} on a finite set $B$ is a set of permutations of $B$ closed under composition. It may easily be verified that such a set contains the identity and is closed under inverse. A permutation subgroup may be identified with a particular type of DSM in which all shops have only singleton sets in their range. The permutation subgroups form a lattice under inclusion whose minimal element contains just the identity and whose maximal element is the symmetric group $S_{|B|}$. As per the Galois connection of the previous section, this lattice classifies the complexities of $\posFO(\mathcal{B})$ -- although we shall see these are relatively uninteresting.

In the lattice of DSMs, the minimal element still contains just $id_B$, but the maximal element contains all shops. However, the lattice of permutation subgroups always appears as a sub-lattice within the lattice of DSMs.

\section{Classification methods}
\label{sec:classification-methods}

We are now in a position to study the interplay between the shes of a structure $\mathcal{B}$ and the complexity of the problem $\mylogic(\mathcal{B})$.

\subsection{Shes inducing lower complexity}
\label{sec:shes-inducing-lower-complexity}

We begin by studying three classes of she, the presence of any of which reduces the complexity of the problem $\mylogic(\mathcal{B})$. Let $\mathcal{B}$ be a finite structure, with distinct elements $b,b'$. We define the following shops from $B$ to $\mathfrak{P}(B) \setminus \{\emptyset\}$.
\[ \forall_b(x):= \left\{ \begin{array}{cl} B & \mbox{if $x=b$} \\ \{x\} & \mbox{otherwise.} \end{array} \right. \]
\[ \exists_b(x):= \{x,b\} \]
\[ \forall_b \exists_{b'}(x):= \left\{ \begin{array}{cl} B & \mbox{if $x=b$} \\ \{b'\} & \mbox{otherwise.} \end{array} \right. \]
We call their classes $\forall$-, $\exists$- and $\forall \exists$-shops, respectively. 
\begin{figure}[h]
\label{fig:first-four}
\hspace{.8cm} 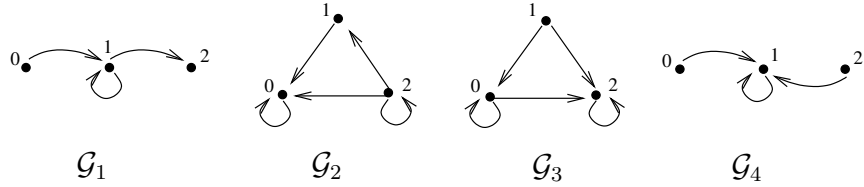
\caption{Sample digraphs admitting $\forall$-, $\exists$- and $\forall \exists$-hyper-operations as shes.}
\end{figure}
In Figure~1, four digraphs $\mathcal{G}_1$--$\mathcal{G}_4$ are drawn. For typographic reasons we will mark-up, e.g., the shop $0 \mapsto \{0,1\}$, $1 \mapsto \{1\}$ and $2 \mapsto \{1,2\}$ as $\she{01}{1}{12}$. It may easily be verified that the DSMs $\shE(\mathcal{G}_1)$--$\shE(\mathcal{G}_4)$ are as follows.
\[
\begin{array}{cccc}
\shE(\mathcal{G}_1) & \shE(\mathcal{G}_2) & \shE(\mathcal{G}_3) & \shE(\mathcal{G}_4) \\
\langle \she{01}{1}{12} \rangle & \langle \she{0}{012}{2} \rangle & \langle \she{2}{012}{2}, \she{0}{01}{2} \rangle & \langle \she{012}{1}{012} \rangle \\
\end{array}
\]
We see that $\mathcal{G}_1$, $\mathcal{G}_2$ and $\mathcal{G}_3$ admit the shes $\exists_1$, $\forall_1$ and $\forall_1 \exists_2$, respectively. $\mathcal{G}_4$ admits each of the shes $\forall_0$, $\forall_2$, $\exists_1$, $\forall_0 \exists_1$ and $\forall_2 \exists_1$.
\begin{remarks}
We have not considered shes $\forall_b \exists_{b}$, defined as above but with $b':=b$. The DSM $\langle \forall_b \exists_{b} \rangle$ is easily seen to contain all shops. It follows that any structure $\mathcal{B}$ that has $\forall_b \exists_{b}$ as a she already has all shes of the form $\forall_{b'} \exists_{b''}$ with $b'\neq b''$.

Note that the DSMs $\langle \forall_b \exists_{b'} \rangle$ and $\langle \{\forall_b, \exists_{b'}\} \rangle = \langle 
\forall_b \circ \exists_{b'} \rangle = \langle 
\exists_{b'} \circ \forall_b \rangle$ do not in general coincide, though the first is always a subset of the following three. Also, we note the identities $\exists_b^{-1}=\forall_b$, $\forall_b^{-1}=\exists_b$ and $(\forall_b \exists_{b'})^{-1}=\forall_{b'} \exists_{b}$.
\end{remarks}
We now give a series of three lemmas, one associated with each of the shops $\forall_b$, $\exists_b$ and $\forall_b \exists_{b'}$. They will ultimately be used in a form of quantifier elimination that will diminish the complexity of $\mylogic(\mathcal{B})$, if $\mathcal{B}$ has one of these as a she.
\begin{lemma}
\label{lem:forall-canon}
Let $\varphi(u,\tuple{v})$ be a formula of \mylogic. Let $\mathcal{B}$ be a finite structure with $\forall_b$ as a she. Then 
\[ \mathcal{B} \models \forall u \ \varphi(u,\tuple{v}) \ \Longleftrightarrow \ \mathcal{B} \models \varphi(b,\tuple{v}). \]
\end{lemma}
\begin{proof}
The forward direction is trivial; we prove the backward. Consider the relation defined by the formula $\varphi(u,\tuple{v})$, where $\tuple{v}:=(v_1,\ldots,v_k)$, of $\mylogic$ on $\mathcal{B}$. By Theorem~\ref{thm:galois-connection}, it is invariant under $\forall_b \in \shE(\mathcal{B})$. For any $x_1,\ldots,x_k \in B$, assume $\mathcal{B} \models \varphi(b,x_1,\ldots,x_k)$. Taking an arbitrary $c \in B$, and noting each $x_i \in \forall_b(x_i)$ and $c \in \forall_b(b)$, we derive $\mathcal{B} \models \varphi(c,x_1,\ldots,x_k)$. The result follows.
\end{proof}
\begin{lemma}
\label{lem:exists-canon}
Let $\varphi(u,\tuple{v})$ be a formula of \mylogic. Let $\mathcal{B}$ be a finite structure with $\exists_b$ as a she. Then 
\[ \mathcal{B} \models \exists u \ \varphi(u,\tuple{v}) \ \Longleftrightarrow \ \mathcal{B} \models \varphi(b,\tuple{v}). \]
\end{lemma}
\begin{proof}
The backward direction is trivial; we prove the forward. Consider the relation defined by the formula $\varphi(u,\tuple{v})$, where $\tuple{v}:=(v_1,\ldots,v_k)$, of $\mylogic$ on $\mathcal{B}$. By Theorem~\ref{thm:galois-connection}, it is invariant under $\exists_b \in \shE(\mathcal{B})$. For any $x_1,\ldots,x_k \in B$, and some $c \in B$, assume $\mathcal{B} \models \varphi(c,x_1,\ldots,x_k)$. Noting each $x_i \in \exists_b(x_i)$ and $b \in \exists_b(c)$, we derive $\mathcal{B} \models \varphi(b,x_1,\ldots,x_k)$. The result follows.
\end{proof}
\begin{lemma}
\label{cor:interpolation}
Let $\varphi(u,\tuple{v})$ be a formula of \mylogic, where the arity of $\tuple{v}$ is $k$. Let $\mathcal{B}$ be a finite structure with $\forall_b \exists_{b'}$ as a she. For all $c \in B$ and $\tuple{x}:=(x_1,\ldots,x_k) \in$ $ \{b,b'\}^k$,
\[ \mathcal{B} \models \varphi(b,\tuple{x}) \ \stackrel{(I)}{\Longrightarrow} \ \mathcal{B} \models \varphi(c,\tuple{x}) \ \stackrel{(II)}{\Longrightarrow} \ \mathcal{B} \models \varphi(b',\tuple{x}). \]
\end{lemma}
\begin{proof}
Consider the relation defined by the formula $\varphi(u,\tuple{v})$ of $\mylogic$ on $\mathcal{B}$. By Theorem~\ref{thm:galois-connection}, it is invariant under $\forall_b \exists_{b'} \in \shE(\mathcal{B})$. Take arbitrary $c \in B$. Noting that $x_i$ is from $\{b,b'\}$, we have $x_i \in \forall_b \exists_{b'}(x_i)$ and $c \in \forall_b \exists_{b'}(b)$. Part $(I)$ follows. Now noting that $b' \in \forall_b \exists_{b'}(c)$, Part $(II)$ follows.
\end{proof}
We are now ready to state how the presence of $\forall$-, $\exists$- or $\forall \exists$-shops as shes of $\mathcal{B}$ can diminish the complexity of $\mylogic(\mathcal{B})$. In each case we proceed by quantifier elimination.
\begin{theorem}
\label{thm:canons}
If $\mathcal{B}$ has a $\forall$-shop as a she then $\mylogic(\mathcal{B})$ is in \NP. If $\mathcal{B}$ has an $\exists$-shop as a she then $\mylogic(\mathcal{B})$ is in \coNP. If $\mathcal{B}$ has a $\forall \exists$-shop as a she then $\mylogic(\mathcal{B})$ is in \Logspace.
\end{theorem}
\begin{proof}
Let $\varphi$ be a sentence of $\mylogic$, and let $\varphi_{[\forall/b]}$ (respectively, $\varphi_{[\exists/b]}$ and $\varphi_{[\forall/b,\exists/b']}$) be $\varphi$ with all universal variables substituted by $b$ (respectively, existential variables substituted by $b$ and universal variables substituted by $b$ and existential variables substituted by $b'$).

If $\mathcal{B}$ has a she $\forall_b$, then consider a sentence $\varphi \in \mylogic$, w.l.o.g. in prenex form. It follows by repeated application of Lemma~\ref{lem:forall-canon} on $\varphi$ -- either from the outermost quantifier in, or from the innermost quantifier out -- that $\mathcal{B} \models \varphi$ iff $\mathcal{B} \models \varphi_{[\forall/b]}$. Similarly, if $\mathcal{B}$ has a she $\exists_{b'}$, then it follows by repeated application of Lemma~\ref{lem:exists-canon} that $\mathcal{B} \models \varphi$ iff $\mathcal{B} \models \varphi_{[\exists/b']}$.

If $\mathcal{B}$ has a she $\forall_b \exists_{b'}$, then, again, assume the sentence $\varphi \in \mylogic$ to be in prenex form. It follows by repeated application of Lemma~\ref{cor:interpolation} -- from the outermost quantifier in -- that $\mathcal{B} \models \varphi$ iff $\mathcal{B} \models \varphi_{[\forall/b,\exists/b']}$. Note that, in this case, one can not move from the innermost quantifier out because this may involve the possibility of free variables taking values from outside the set $\{b,b'\}$. 
The result now follows since evaluating $\varphi_{[\forall/b,\exists/b']}$ on $\mathcal{B}$ is equivalent to a boolean sentence value problem, known to be in \Logspace\ \cite{NLynch}.
\end{proof}
Returning to the examples of Figure~1, we see that $\mylogic(\mathcal{G}_1)$ is in \coNP, $\{ \exists, \forall, \wedge, \vee \}$-$\FO(\mathcal{G}_2)$ is in \NP\ and both $\mylogic(\mathcal{G}_3)$ and $\mylogic(\mathcal{G}_4)$ are in \Logspace.  

\subsection{Extending the method}

Call a shop $f:B \rightarrow \mathfrak{P}(B) \setminus \{\emptyset\}$ an \emph{$\mathsf{A}$-shop} if there exists $b \in B$ s.t. $f(b)=B$. Call $f$ an \emph{$\mathsf{E}$-shop} if there exists $b \in B$ s.t. $b \in f(x)$, for all $x \in B$. The generalisation of $\forall$- and $\exists$-shops is clear. Let $f^r$ denote $f$ composed with itself $r$ times. It is clear that, if $f$ is an $\mathsf{A}$-shop (resp., an $\mathsf{E}$-shop) as just defined, then $f^r(b)=B$ (resp., $b \in f^r(x)$, for all $x \in B$), for all $r$. With an arbitrary shop $f$ on $B$, we may associate the digraph $\mathscr{G}_f$ on $B$ in which there is an edge $(x,y)$ if $f(x) \ni y$. In a digraph, a \emph{source} is a vertex of in-degree zero and a \emph{sink} is a vertex of out-degree zero (a self-loop is neither). The condition of totality ensures $\mathscr{G}_f$ has no sinks and the condition of surjectivity ensures $\mathscr{G}_f$ has no sources.
\begin{lemma}
\label{lem:A-hyperop}
Let $f$ be an $\mathsf{A}$-shop on a set $B$, with $|B|\geq 2$. Then $\langle f \rangle$ contains an $\mathsf{A}$-shop $g$ with a tripartition $\{b\}; B'; B''$ of $B$ ($B'$ non-empty) s.t.
\begin{itemize}
\item $g(b)=B$
\item for all $x \in B'$, $g(x)=\{x\}$
\item for all $x \in B''$, there exists $y \in B'$, $g(x)=\{y\}$.
\end{itemize}
\end{lemma}
\begin{proof}
Let $b$ be s.t. $f(b)=B$. It is possible that there is $x \in B$ s.t. $f(x)=\{b\}$. However, by considering paths in $\mathscr{G}_f$, it is easy to see that
\[
\begin{array}{lc}
\mbox{(\ref{lem:A-hyperop} *)} & \mbox{ for no $x$ does } f^{|B|}(x)=\{b\}.
\end{array}
\] 
Consider now $\mathscr{G}'_{f^{|B|}}$ to be the graph $\mathscr{G}_{f^{|B|}}$ with the vertex $b$ removed. Owing to (\ref{lem:A-hyperop} *), $\mathscr{G}'_{f^{|B|}}$ will still have no sinks, but it may now have sources. Build $\mathscr{G}''_{f^{|B|}}$  from $\mathscr{G}'_{f^{|B|}}$ by recursively removing sources from $\mathscr{G}'_{f^{|B|}}$ until none is left. Let us say this takes $d$ steps. $\mathscr{G}''_{f^{|B|}}$ is therefore the disjoint union of strongly-connected components. For each of its strongly-connected components $C_1,\ldots,C_k$ pick a cycle (not necessarily Hamiltonian) that visits each vertex in the component at least once. Let the lengths of these cycles be $c_1,\ldots,c_k$ and let $c$ be the least common multiple of $\{c_1,\ldots,c_k,d\}$. It is not hard to see that some sub-shop $g$ of $(f^{|B|})^c$ has the desired properties, with $B'$ being those vertices that remain in $\mathscr{G}''_{f^{|B|}}$.
\end{proof}
\begin{lemma}
\label{lem:E-hyperop}
Let $f$ be an $\mathsf{E}$-shop on a set $B$, with $|B|\geq 2$, where $b \in f(x)$ for all $x$. Then $\langle f \rangle$ contains an $\mathsf{E}$-shop $g$ with a bipartition $B'; B''$ of $B$ ($B'$ non-empty) s.t.
\begin{itemize}
\item for all $x \in B'$, $g(x) \supseteq \{x,b\}$
\item for all $x \in B''$, $g(x) \supseteq \{b\}$
\item for all $y \in B$, exists $x \in B'$, $y \in g(x)$.
\end{itemize}
\end{lemma}
\begin{proof}
It is possible that there is $x \in B$ s.t. 
\[ x \in f(b) \mbox{ but for all $y \in B\setminus \{b\}$, $x \notin f(y)$}.\]
However, by considering paths in $\mathscr{G}_f$, it is easy to see that
\[
\begin{array}{lc}
\mbox{(\ref{lem:E-hyperop} *)} & \mbox{ there is no $x \in B$ s.t. $x \in f^{|B|}(b)$ but for all $y \in B\setminus \{b\}$, $x \notin f^{|B|}(y)$}.
\end{array}
\] 
Consider now $\mathscr{G}'_{f^{|B|}}$ to be the graph $\mathscr{G}_{f^{|B|}}$ with the vertex $b$ removed. Owing to (\ref{lem:E-hyperop} *), $\mathscr{G}'_{f^{|B|}}$ will still have no sources, but it may now have sinks. Build $\mathscr{G}''_{f^{|B|}}$  from $\mathscr{G}'_{f^{|B|}}$ by recursively removing sinks from $\mathscr{G}'_{f^{|B|}}$ until none is left. Let us say this takes $d$ steps. $\mathscr{G}''_{f^{|B|}}$ is therefore the disjoint union of strongly-connected components. For each of its strongly-connected components $C_1,\ldots,C_k$ pick a cycle (not necessarily Hamiltonian) that visits each vertex in the component at least once. Let the lengths of these cycles be $c_1,\ldots,c_k$ and let $c$ be the least common multiple of $\{c_1,\ldots,c_k,d\}$. It is not hard to see that some sub-shop $g$ of $(f^{|B|})^c$ has the desired properties, with $B'$ being those vertices that remain in $\mathscr{G}''_{f^{|B|}}$.
\end{proof}
\noindent We give the following examples of shops $g$ of the given forms.
\[
\begin{array}{cc}
\mbox{Lemma~\ref{lem:A-hyperop}} & \mbox{Lemma~\ref{lem:E-hyperop}} \\
\shefour{0123}{1}{1}{3} & \shefour{0}{0}{012}{03}  \\
\end{array}
\]
\begin{lemma}
\label{cor:interpolation-A}
Let $\varphi(u,\tuple{v})$ be a formula of \mylogic, where the arity of $\tuple{v}$ is $k$. Let $\mathcal{B}$ be a finite structure with an $\mathsf{A}$-shop $g$, satisfying the conditions of Lemma~\ref{lem:A-hyperop}, as a she. For all $\tuple{x}:=(x_1,\ldots,x_k) \in$ $(B' \cup \{b\})^k$,
\[ \mathcal{B} \models \varphi(b,\tuple{x}) \ \stackrel{(I)}{\Longrightarrow} \ \mathcal{B} \models \forall u \ \varphi(u,\tuple{x}) \]
\[ \mathcal{B} \models \exists u \ \varphi(u,\tuple{x}) \ \stackrel{(II)}{\Longrightarrow} \ \mathcal{B} \models \exists u \in B' \ \varphi(u,\tuple{x}). \]
\end{lemma}
\begin{proof}
Consider the relation defined by the formula $\varphi(u,\tuple{v})$ of $\mylogic$ on $\mathcal{B}$. By Theorem~\ref{thm:galois-connection}, it is invariant under $g \in \shE(\mathcal{B})$. Take arbitrary $u \in B$. Noting that $x_i$ is from $B' \cup \{b\}$, we have $x_i \in g(x_i)$ and $u \in g(b)$. Part $(I)$ follows. Now noting that, for each $u \in B$ there is some $u' \in B'$ s.t. $u' \in g(u)$, Part $(II)$ follows.
\end{proof}
\begin{lemma}
\label{cor:interpolation-E}
Let $\varphi(u,\tuple{v})$ be a formula of \mylogic, where the arity of $\tuple{v}$ is $k$. Let $\mathcal{B}$ be a finite structure with an $\mathsf{E}$-shop $g$, satisfying the conditions of Lemma~\ref{lem:E-hyperop}, as a she. For all $\tuple{x}:=(x_1,\ldots,x_k) \in$ $(B' \cup \{b\})^k$,
\[ \mathcal{B} \models \forall u \in B' \ \varphi(u,\tuple{x}) \ \stackrel{(I)}{\Longrightarrow} \ \mathcal{B} \models  \forall u \ \varphi(c,\tuple{x}) \]
\[ \mathcal{B} \models \exists u \ \varphi(u,\tuple{x}) \ \stackrel{(II)}{\Longrightarrow} \ \mathcal{B} \models \varphi(b,\tuple{x}). \]
\end{lemma}
\begin{proof}
Consider the relation defined by the formula $\varphi(u,\tuple{v})$ of $\mylogic$ on $\mathcal{B}$. By Theorem~\ref{thm:galois-connection}, it is invariant under $g \in \shE(\mathcal{B})$. Take arbitrary $u \in B$. Noting that $x_i$ is from $B' \cup \{b\}$, and $x_i \in g(x_i)$, and that there is some $u' \in B'$ s.t. $u \in g(u')$, Part $(I)$ follows. Now noting that, for each $u \in B$ we have $b \in g(u)$, Part $(II)$ follows.
\end{proof}
\begin{theorem}
\label{thm:canons2}
If $\mathcal{B}$ has an $\mathsf{A}$-shop as a she then $\mylogic(\mathcal{B})$ is in \NP. If $\mathcal{B}$ has an $\mathsf{E}$-shop as a she then $\mylogic(\mathcal{B})$ is in \coNP. If $\mathcal{B}$ has a both an $\mathsf{A}$-shop and an $\mathsf{E}$-shop as a she then $\mylogic(\mathcal{B})$ is in \Logspace.
\end{theorem}
\begin{proof}
Let $\varphi$ be a sentence of $\mylogic$, and let $\varphi_{[\forall/b,\exists/B']}$ (respectively, $\varphi_{[\exists/b,\forall/B']}$) be $\varphi$ with all universal variables substituted by $b$ and existential variables restricted to $B'$ (respectively, existential variables substituted by $b$ and universal variables restricted to $B'$).

If $\mathcal{B}$ has an $\mathsf{A}$-shop $f$ as a she, then let $g \in \langle f \rangle$ be as in Lemma~\ref{lem:A-hyperop}. Consider a sentence $\varphi \in \mylogic$, w.l.o.g. in prenex form. It follows by repeated application of Lemma~\ref{cor:interpolation-A} on $\varphi$ -- from the outermost quantifier in -- that $\mathcal{B} \models \varphi$ iff $\mathcal{B} \models \varphi_{[\forall/b,\exists/B']}$. Similarly, if $\mathcal{B}$ has an $\mathsf{E}$-shop as a she then it follows by repeated application of Lemma~\ref{cor:interpolation-E} that $\mathcal{B} \models \varphi$ iff $\mathcal{B} \models \varphi_{[\exists/b,\forall/B']}$.

If $\mathcal{B}$ has both an $\mathsf{A}$-shop $f_A$ and an $\mathsf{E}$-shop $f_E$ as a she, then it follows that their composition $f_A \circ f_E$ (also $f_E \circ f_A$) is a she and an $\exists \forall$-shop. The result follows from Theorem~\ref{thm:canons}. 
\end{proof}

\subsection{Reduction to simpler cases}

For a shop $f$, we recall the associated digraph $\mathscr{G}_f$ defined in the previous section. We say that $f$ is an equivalence relation if $\mathscr{G}_f$ is the digraph of an equivalence relation ($f$ maps each element to its equivalence class). For a structure $\mathcal{B}$ and an equivalence relation $f$ on $B$, we define the structure $\mathcal{B}_{/f}$ as follows. The elements of $\mathcal{B}_{/f}$ are the equivalence classes of $f$ and the relation $R^{\mathcal{B}_{/f}}(\tilde{b}_1,\ldots,\tilde{b}_r)$ holds if for some representatives $b_1,\ldots,b_r$ of the classes $\tilde{b}_1,\ldots,\tilde{b}_r$ the relation $R^{\mathcal{B}_{/f}}(b_1,\ldots,b_r)$ holds.
\begin{lemma}
\label{lem:equiv-classes}
Let $\mathcal{B}$ be a structure with an equivalence relation $f$ as a she. Then $\mylogic(\mathcal{B})=\mylogic(\mathcal{B}_{/f})$.
\end{lemma}
\begin{proof}
In fact, it is easy to see that $\mathcal{B}$ and $\mathcal{B}_{/f}$ agree on all sentences of equality-free FO logic. This is because the she $f$ guarantees that all elements in an equivalence class partake in exactly the same relations as one another. Indeed, there is a strong homomorphism from $\mathcal{B}$ to $\mathcal{B}_{/f}$ (for more details, see, e.g., the Homomorphism Theorem in \cite{Enderton}).
\end{proof}

\subsection{Down-she-monoids of high complexity}

\begin{lemma}
\label{lem:permutation-subgroups}
Let $\mathcal{B}$, with $|B| \geq 2$, be a structure s.t. $\shE(\mathcal{B})$ is a permutation subgroup. Then $\mylogic(\mathcal{B})$ is \Pspace-complete.
\end{lemma}
\begin{proof}
Let $\mathcal{B}_{NAE}$ be the structure on $B$ with a single ternary relation $R_{NAE}:= B^3 \setminus \{(b,b,b) : b \in B\}$. $\mylogic(\mathcal{B}_{NAE})$ is a generalisation of the problem $\QCSP(\mathcal{B}_{NAE})$, well-known to be \Pspace-complete (in the case $|B|=2$, this is \emph{quantified not-all-equal $3$-satisfiability}, see, e.g., \cite{ComputationalComplexity}). $\shE(\mathcal{B}_{NAE})$ is the symmetric group $S_{|B|}$. The statement of the theorem now follows from Theorem~\ref{thm:she-reduction}, since $\shE(\mathcal{B}) \subseteq \shE(\mathcal{B}_{NAE})$.
\end{proof}
\begin{corollary}
For all $\mathcal{B}$ s.t. $|B| \geq 2$, $\posFO(\mathcal{B})$ is \Pspace-complete.
\end{corollary}
\begin{proof}
$\posFO(\mathcal{B})$ may be rephrased as the problem $\mylogic(\mathcal{B}')$, where $\mathcal{B}'$ is the structure $\mathcal{B}$ expanded with the graph of equality. Owing to the presence of the graph of equality, $\shE(\mathcal{B}')$ must be a permutation subgroup, and the result follows from the previous lemma.  
\end{proof}
\noindent The following is a generalisation of Lemma~\ref{lem:permutation-subgroups}.
\begin{lemma}
\label{lem:collapse-permutation-subgroups}
Let $\mathcal{B}$ be a structure whose universe admits the partition $B_1,\ldots,B_l$ ($l \geq 2$). If all shes of $\mathcal{B}$ are sub-hyper-operations of some $f$ of the form $f(x) := B_i$ iff $x \in B_{\pi(i)}$, for $\pi$ a permutation on the set $\{1,\ldots,l\}$, then $\mylogic(\mathcal{B})$ is \Pspace-complete.
\end{lemma}
\begin{proof}
Let $\mathcal{K}_{|B_1|,\ldots,|B_l|}$ be the complete $l$-partite graph with partitions of size $|B_1|,\ldots,|B_l|$. It may easily be verified that the $\shE(\mathcal{B}) \subseteq \shE(\mathcal{K}_{|B_1|,\ldots,|B_l|})$. $\mathcal{K}_{|B_1|,\ldots,|B_l|}$ clearly has the equivalence relation $g$ that maps each element in $B_i$ to $B_i$ ($1 \leq i \leq l$) as a she. Hence, $\mylogic(\mathcal{K}_{|B_1|,\ldots,|B_l|})=$ $\mylogic(\mathcal{K}_l)$ by Lemma~\ref{lem:equiv-classes}, and the result follows from Theorem~\ref{thm:she-reduction} and Lemma~\ref{lem:permutation-subgroups}.
\end{proof}
\begin{figure}
\label{fig:second-four}
\hspace{.8cm} 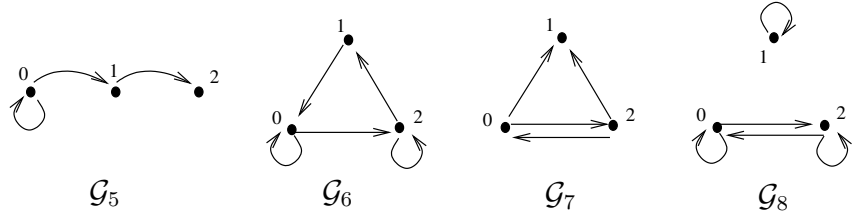
\caption{Further sample digraphs.}
\end{figure}
In Figure~2, four more digraphs $\mathcal{G}_5$--$\mathcal{G}_8$ are drawn. It may easily be verified that $\shE(\mathcal{G}_5)$--$\shE(\mathcal{G}_8)$ are as follows. 
\[
\begin{array}{cccc}
\shE(\mathcal{G}_5) & \shE(\mathcal{G}_6) & \shE(\mathcal{G}_7) & \shE(\mathcal{G}_8) \\
\langle \she{0}{1}{2} \rangle & \langle \she{0}{1}{2} \rangle & \langle \she{0}{2}{1} \rangle & \langle \she{02}{1}{02} \rangle \\
\end{array}
\]
It follows from Lemmas~\ref{lem:permutation-subgroups} and \ref{lem:collapse-permutation-subgroups} that each of $\mylogic(\mathcal{G}_5)$, \ldots, $\{ \exists, \forall, \wedge, \vee\}$-$\FO(\mathcal{G}_8)$ is \Pspace-complete.

\section{Classification results}
\label{sec:classification-results}

We are now in a position to use the methods of the previous section to classify the complexities of $\mylogic(\mathcal{B})$ as $\mathcal{B}$ ranges over, firstly, boolean structures, and then structures of size three.

\subsection{The boolean case}

We consider the case $|B|=2$, with the normalised domain $B:=\{0,1\}$. 
It may easily be verified that there are five DSMs in this case, depicted as a lattice in Figure~3. The two elements of this lattice that represent the two subgroups of $S_2$ are drawn in the middle and bottom.
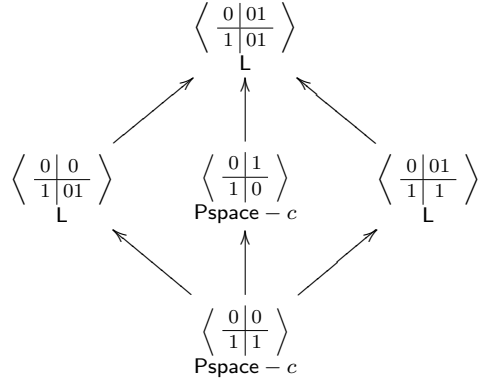
\begin{figure}[h]
\label{fig:boolean-she-lattice}
\[
\xymatrix{
& \stackrel{\mbox{\color{black}{$\left\langle \shee{01}{01} \right\rangle$}}}{\Logspace} & \\
\stackrel{\mbox{\color{black}{$\left\langle \shee{0}{01} \right\rangle$}}}{\Logspace} \ar[ur] & \stackrel{\left\langle \shee{1}{0} \right\rangle}{\Pspace -c} \ar[u] & \stackrel{\mbox{\color{black}{$\left\langle \shee{01}{1} \right\rangle$}}}{\Logspace} \ar[ul] \\
& \stackrel{\left\langle \shee{0}{1} \right\rangle}{\Pspace -c} \ar[ul] \ar[u] \ar[ur] & \\
}
\]
\caption{The boolean lattice of DSMs with their associated complexity.}
\end{figure}
\begin{theorem}[Dichotomy]
Let $\mathcal{B}$ be a boolean structure. 
\begin{itemize}
\item[I.] If either $\forall_0 \exists_{1}$ or $\forall_1 \exists_{0}$ (i.e., $\shee{01}{1}$ or $\shee{0}{01}$) is a she of $\mathcal{B}$, then $\mylogic(\mathcal{B})$ is in \Logspace.
\item[II.] Otherwise, $\mylogic(\mathcal{B})$ is \Pspace-complete.
\end{itemize}
\end{theorem}
\begin{proof}
$\shE(\mathcal{B})$ must be one of the five DSMs depicted in Figure~3. If $\shE(\mathcal{B})$ contains one of $\forall_0 \exists_{1}$ or $\forall_1 \exists_{0}$, then \Logspace\ membership follows from Theorem~\ref{thm:canons}. Otherwise $\shE(\mathcal{B})$ is either $\langle \shee{0}{1} \rangle$ or $\langle \shee{1}{0} \rangle$; in both cases the hardness result follows from Lemma~\ref{lem:permutation-subgroups}.
\end{proof}
\begin{remark}
In the boolean case, $\langle \forall_1 \exists_0 \rangle = \langle \{ \forall_1, \exists_0 \}\rangle$ and $\langle \forall_0 \exists_1 \rangle = \langle \{ \forall_0, \exists_1 \}\rangle$.
\end{remark}

\subsection{The three-element case}

\begin{figure}
\label{fig:K2uplusK1-and-complement}
\hspace{3.6cm} 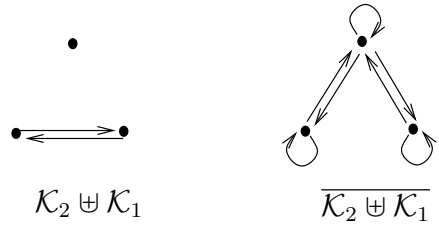
\caption{The digraphs involved in Classes II and III of Theorem~\ref{thm:tetrachotomy}.}
\end{figure}

We consider the case $|B|=3$, with the normalised domain $B:=\{0,1,2\}$. We will move straight to the classification theorem.
\begin{theorem}[Tetrachotomy]
\label{thm:tetrachotomy}
Let $\mathcal{B}$ be a three-element structure. 
\begin{itemize}
\item[I.] If $\shE(\mathcal{A})$ contains both an $\mathsf{A}$-shop and an $\mathsf{E}$-shop, then $\mylogic(\mathcal{B})$ is in \Logspace.
\item[II.] If $\shE(\mathcal{A})$ contains an $\mathsf{A}$-shop but no $\mathsf{E}$-shop, then $\mylogic(\mathcal{B})$ is \NP-complete.
\item[III.] If $\shE(\mathcal{A})$ contains an $\mathsf{E}$-shop but no $\mathsf{A}$-shop, then $\mylogic(\mathcal{B})$ is \coNP-complete.
\item[IV.] If $\shE(\mathcal{A})$ contains neither an $\mathsf{A}$-shop nor an $\mathsf{E}$-shop, then $\mylogic(\mathcal{B})$ is \Pspace-complete.
\end{itemize}
\end{theorem}
\begin{proof} \

\textbf{I. $\shE(\mathcal{A})$ contains both an $\mathsf{A}$-shop and an $\mathsf{E}$-shop}.
If $\shE(\mathcal{A})$ contains both an $\mathsf{A}$-shop $f$ and an $\mathsf{E}$-shop $g$, then both $f \circ g$ and $g \circ f$ are $\forall \exists$-shops, and it follows from Theorem~\ref{thm:canons} that $\mylogic(\mathcal{A})$ is in \Logspace.

\

\textbf{II. $\shE(\mathcal{A})$ contains an $\mathsf{A}$-shop but no $\mathsf{E}$-shop}.
Removing a symmetry, and without loss of generality, we assume that $\shE(\mathcal{A})$ contains a shop $f$ s.t. $f(0)=\{0,1,2\}$. If either $1$ or $2$ were in both $f(1)$ and $f(2)$ then $f$ would be a $\forall \exists$-shop. If either $f(1)$ or $f(2)$ were $\{1,2\}$  then either $f$ would be a $\forall \exists$-shop or the other of $f(1)$ or $f(2)$ would contain $0$. And, if either of $f(1)$ or $f(2)$ contained $0$, then $f^2$ would be a $\forall \exists$-shop. It follows that $f$ must be either $\she{012}{1}{2}$ or $\she{012}{2}{1}$, and that $\she{012}{1}{2}$ necessarily appears in $\shE(\mathcal{A})$. 

Which other shops may also appear in $\shE(\mathcal{A})$? Again no $g$ with any of the properties 
\begin{enumerate}
\item either $1$ or $2$ are in both $g(1)$ and $g(2)$,
\item either $g(1)$ or $g(2)$ are $\{1,2\}$,
\item either $g(1)$ or $g(2)$ contain $0$,
\end{enumerate}
since then $g \circ \she{012}{1}{2}$ would be in one of the forbidden as in the previous paragraph. It follows that only shops among $\langle \she{012}{2}{1}\rangle$ may also be in $\shE(\mathcal{A})$. Remembering symmetries, it follows that $\shE(\mathcal{A})$ is one of the six DSMs, $\langle \she{012}{1}{2} \rangle$, $\langle \she{012}{2}{1}\rangle$, $\langle \she{1}{012}{2}\rangle$, $\langle \she{2}{012}{1}\rangle$, $\langle \she{1}{2}{012}\rangle$ and $\langle \she{2}{1}{012}\rangle$. Membership of \NP\ follows from either of Theorems~\ref{thm:canons} or \ref{thm:canons2}. For \NP-hardness, consider the disjoint union $\mathcal{K}_2 \uplus \mathcal{K}_1$ of the antireflexive $2$- and $1$-cliques, as drawn in Figure~\ref{fig:K2uplusK1-and-complement}. $\shE(\mathcal{K}_2 \uplus \mathcal{K}_1)$ is either $\langle \she{012}{2}{1} \rangle$, $\langle \she{2}{012}{0} \rangle$ or $\langle \she{1}{0}{012} \rangle$, depending on the vertex labelling. $\{ \exists, \wedge, \vee\}$-FO$(\mathcal{K}_2)$ is \NP-complete (by reduction from \emph{$3$-not-all-equal satisfiablity}, set $R_{NAE}(u,v,w):=E(u,v) \vee E(v,w)$), and $\mathcal{K}_2 \uplus \mathcal{K}_1$ agrees with $\mathcal{K}_2$ on all sentences of $\{ \exists, \wedge, \vee\}$-FO (see \cite{CiE2008}). It follows that $\{ \exists, \wedge, \vee\}$-FO$(\mathcal{K}_2 \uplus K_1)$ is \NP-complete and that $\mylogic(\mathcal{K}_2 \uplus K_1)$ is \NP-hard. The result for \NP-hardness now follows from Theorem~\ref{thm:she-reduction}, since $\shE(\mathcal{B}) \subseteq \shE(\mathcal{K}_2 \uplus K_1)$, for one of the three vertex labellings.

\

\textbf{III. $\shE(\mathcal{A})$ contains an $\mathsf{E}$-shop but no $\mathsf{A}$-shop}.
Removing a symmetry, and without loss of generality, we assume that $\shE(\mathcal{A})$ contains a shop $f$ s.t. $0 \in f(0),f(1),f(2)$. If either $f(1)$ or $f(2)$ also contained $\{1,2\}$, then $f$ would be a $\forall \exists$-shop. If either $1$ or $2$ were in both $f(1)$ and $f(2)$ then either $f$ would be a $\forall \exists$-shop or the other of $1$ or $2$ would be in $f(0)$. And, if $f(0)$ contained either $1$ or $2$ (e.g. $f$ looked like $\she{01\ldots}{0\ldots}{0\ldots}$), then $f^2$ would be a $\forall \exists$-shop.  It follows that $f$ must be either $\she{0}{01}{02}$ or $\she{0}{02}{01}$, and that $\she{0}{01}{02}$ necessarily appears in $\shE(\mathcal{A})$. 

Which other shops may also appear in $\shE(\mathcal{A})$? Again no $g$ with any of the properties 
\begin{enumerate}
\item either $g(1)$ or $g(2)$ contains $\{1,2\}$,
\item either $1$ or $2$ in both $g(1)$ and $g(2)$,
\item $g(0)$ contains either $1$ or $2$,
\end{enumerate}
since then $g \circ \she{0}{01}{02}$ would be forbidden as in the previous paragraph. It follows that only shops among $\langle \she{0}{02}{01}\rangle$ may also be in $\shE(\mathcal{A})$. Remembering symmetries, it follows that $\shE(\mathcal{A})$ is one of the six DSMs, $\langle \she{0}{02}{01}\rangle$, $\langle \she{0}{01}{02}\rangle$, $\langle \she{12}{1}{01}\rangle$, $\langle \she{01}{1}{12}\rangle$, $\langle \she{12}{02}{2}\rangle$ and $\langle \she{02}{12}{2}\rangle$. Membership of \coNP\ follows from either of Theorems~\ref{thm:canons} or \ref{thm:canons2}. For \coNP-hardness, consider now the complement graph $\overline{\mathcal{K}_2 \uplus \mathcal{K}_1}$ (for a graph $\mathcal{G}$, define its complement $\overline{\mathcal{G}}$ over the same vertex set to have the complementary edge set -- i.e. $\overline{\mathcal{G}} \models E(x,y)$ iff $\mathcal{G} \notmodels E(x,y)$). It is a simple application of de Morgan duality that $\mylogic(\mathcal{G})$ is in \NP\ (resp., is \NP-complete) iff $\mylogic(\overline{\mathcal{G}})$ is in \coNP\ (resp., is \coNP-complete) -- see \cite{CiE2008}. A similar argument to that for Class II, but with $\overline{\mathcal{K}_2 \uplus \mathcal{K}_1}$ yields the \coNP-hardness result for Class III. 

\

\textbf{IV. $\shE(\mathcal{A})$ contains neither an $\mathsf{E}$-shop nor an $\mathsf{A}$-shop}.
If $\shE(\mathcal{A})$ is not a a sub-DSM of the DSM associated with the symmetric group $S_3=\langle \she{1}{2}{0},\she{0}{2}{1}\rangle$, then we may assume that $\shE(\mathcal{A})$ contains a shop $f$ where some element is mapped to exactly two elements. There are two possibilities, either the element is included among the two it is mapped to, or it is not. Without loss of generality, we consider these two cases separately as follows.

\textit{$f$ is of the form $\she{01}{\ldots}{\ldots}$}.
$2$ must appear somewhere among $f(0),f(1)$ and $f(2)$, but may not appear among $f(0)$ or $f(1)$ as then $f^2$ would be an $\mathsf{A}$-shop. It follows that $2 \in f(2)$, whereupon neither may $0$ be in $f(2)$ (as then $f^2$ would be an $\mathsf{A}$-shop) nor may $1$ be in $f(2)$ (as then either $f$ or $f^2$ would be an $\mathsf{E}$-shop -- depending on which of $0,1 \in f(1)$). It follows that $f(2)=\{2\}$. In fact, all of the remaining possibilities are valid, i.e. $f$ may be any of $\she{01}{0}{2}$, $\she{01}{1}{2}$ and $\she{01}{01}{2}$.  

\textit{$f$ is of the form $\she{12}{\ldots}{\ldots}$}.
 W.l.o.g. we may assume that $0 \in f(1)$, whereupon $1 \notin f(1)$ for otherwise $f^2$ would be an $\mathsf{A}$-shop. It follows that $f(1)=$ either $\{0\}$ or $\{0,2\}$. In fact the second of these is not possible as, if it were so, then: if $2 \in f(2)$, $f$ would be an $\mathsf{E}$-shop, and if $0$ or $1 \in f(2)$, $f^2$ would be an $\mathsf{A}$-shop. Thus, we have $f(1)=\{0\}$. If $1 \in f(2)$ then $f^3$ would be an $\mathsf{A}$-shop and if $2 \in f(2)$ then $f^2$ would be an $\mathsf{E}$-shop. It follows that $f$ is $\she{12}{0}{0}$.

Recalling symmetries, it follows that $f$ is some shop from among $\langle \she{1}{2}{0},\she{0}{2}{1}\rangle$, $\langle \she{12}{0}{0} \rangle$, $\langle \she{1}{02}{1} \rangle$ and $\langle \she{2}{2}{01} \rangle$.
We have in fact demonstrated that if $\shE(\mathcal{A})$ contains neither an $\mathsf{E}$-shop nor an $\mathsf{A}$-shop then all shops of $\shE(\mathcal{A})$ are among the given list. There are several DSMs that can be formed in this manner but we will demonstrate that all of them are sub-DSMs of one of $\langle \she{1}{2}{0},\she{0}{2}{1}\rangle$, $\langle \she{12}{0}{0} \rangle$, $\langle \she{1}{02}{1} \rangle$ and $\langle \she{2}{2}{01} \rangle$.

Arguing via symmetries, this is clear when we consider that $\she{0}{1}{12}$ can not combine with any element of $S_3=\langle \she{1}{2}{0},\she{0}{2}{1}\rangle$ apart from $\she{0}{2}{1}$ (and the identity). For  $\she{0}{1}{12} \circ \she{1}{2}{0} \circ \she{0}{1}{12}$ is an $\mathsf{A}$-shop; $\she{0}{1}{12} \circ \she{1}{0}{2} \circ \she{0}{1}{12}$ is an $\mathsf{A}$-shop; and $\she{0}{1}{12} \circ \she{2}{1}{0} \circ \she{0}{1}{12}$ is an $\mathsf{E}$-shop. 

\Pspace-hardness when $\shE(\mathcal{A}) \subseteq \langle \she{1}{2}{0},\she{0}{2}{1}\rangle$ follows from Lemma~\ref{lem:permutation-subgroups}, while \Pspace-hardness when $\shE(\mathcal{A}) \subseteq 
$ any of $\langle \she{12}{0}{0} \rangle$, $\langle \she{1}{02}{1} \rangle$ follows from Lemma~\ref{lem:collapse-permutation-subgroups}.
\end{proof}
\noindent Casting our mind back to the digraphs $\mathcal{G}_1$ and $\mathcal{G}_2$ of Figure~1, we can read from the previous theorem that $\mylogic(\mathcal{G}_1)$ and $\mylogic(\mathcal{G}_2)$ are \coNP-complete and \NP-complete, respectively.

\section{Final remarks}
\label{sec:final-remarks}

We have introduced the class of problems $\mylogic(\mathcal{B})$ as well as an algebraic framework in which to study their complexity. We hope that we have adequately demonstrated that this class of problems displays complexity-theoretic richness, while not being too resistant to full classification in simple cases. The algebraic method used in our classification for the three-element case gives simple explanation where there previously was none -- if one were to look at the examples of Figures~1 and 2, there is little obvious in their immediate structure that betrays their position in the classification.

We note that our positive algorithms, for membership of \NP, \coNP\ and -- especially -- \Logspace, are uniform, and are based on simple quantifier elimination. Perhaps it is to be hoped that a full classification for the problems $\mylogic(\mathcal{B})$ would make use only of versions of quantifier elimination. In any case, we conjecture that the tetrachotomy of Theorem~\ref{thm:tetrachotomy} extends to all structures $\mathcal{B}$; though we know we would need more sophisticated classes of shes than those of Section~\ref{sec:shes-inducing-lower-complexity} to prove this. 
\begin{conjecture}[Tetrachotomy]
Let $\mathcal{B}$ be any structure. 
\begin{itemize}
\item[I.] If $\shE(\mathcal{A})$ contains both an $\mathsf{A}$-shop and an $\mathsf{E}$-shop, then $\mylogic(\mathcal{B})$ is in \Logspace.
\item[II.] If $\shE(\mathcal{A})$ contains an $\mathsf{A}$-shop but no $\mathsf{E}$-shop, then $\mylogic(\mathcal{B})$ is \NP-complete.
\item[III.] If $\shE(\mathcal{A})$ contains an $\mathsf{E}$-shop but no $\mathsf{A}$-shop, then $\mylogic(\mathcal{B})$ is \coNP-complete.
\item[IV.] If $\shE(\mathcal{A})$ contains neither an $\mathsf{A}$-shop nor an $\mathsf{E}$-shop, then $\mylogic(\mathcal{B})$ is \Pspace-complete.
\end{itemize}
\end{conjecture}
We note also that, unlike the situation with clones and the \CSP, the down-she-monoids associated with a finite domain are always finite. This means that their lattice should be effectively computable for low domain sizes like four or five.

\bibliographystyle{acm}
\bibliography{she}

\end{document}